\documentclass[nofootinbib,preprint,superscriptaddress,showkeys]{revtex4-2}

\usepackage{amsmath,graphicx,hyperref,color,subfig}

\allowdisplaybreaks

\newcommand{\mc}{\mathcal}
\newcommand{\mr}{\mathrm}

\newcommand{\mQ}{\mathcal{Q}}

\newcommand{\be}{\begin{equation}} 
\newcommand{\ee}{\end{equation}} 
\newcommand{\bea}{\begin{eqnarray}} 
\newcommand{\eea}{\end{eqnarray}}

\newcommand{\dg}{\dagger}
\newcommand{\n}{\overline{n}}

\newcommand{\bl}[1]{{\bf{#1}}}

\newcommand{\nnb}{\nonumber} 

\newcommand{\as}{\alpha_s}

\begin{document}



\title{Dijet invariant mass distribution near threshold}

\def\Seoultech{Institute of Convergence Fundamental Studies and School of Natural Sciences, Seoul National University of Science and Technology, Seoul 01811, Korea}

\author{Chul Kim}
\email[E-mail:]{chul@seoultech.ac.kr}
\affiliation{\Seoultech} 

\author{Taehyun Kwon}
\email[E-mail:]{thteddyk@g.seoultech.ac.kr}
\affiliation{\Seoultech}

\begin{abstract} \vspace{0.1cm}\baselineskip 3.5ex 
In this paper, using soft-collinear effective theory we study the invariant mass distribution for dijet production in $e^+e^-$-annihilation. Near threshold, where the dijet takes most of the energy, there arise the large threshold logarithms, which are sensitive to soft gluon radiations. To systematically resum the logarithms, we factorize the scattering cross section into the hard, the collinear, and the soft parts. And we additionally factorize the original soft part into the global soft function and the two collinear-soft functions, where the latter can be combined with the collinear parts to form the fragmentation functions to jet (FFJs). The factorization theorem derived here can be easily applicable to other processes near threshold. Using the factorized result, we show the resummed result for the dijet invariant mass to the accuracy of next-to-leading logarithms. We have also obtained the result in the case of the heavy quark dijet and compared it with the case of the light quark. 
\end{abstract}


\maketitle 
\newpage


\section{Introduction}

Studying soft and collinear physics is important for understanding the physics of high energy processes at a lower scale.  
By properly extracting soft and collinear parts from the full process we correctly obtain the hard part and its renormalization behavior. Then, through renormalization group (RG) evolution, we can systematically resum large logarithms between the hard and the lower scales. For this purpose, soft-collinear effective theory~(SCET)~\cite{Bauer:2000ew,Bauer:2000yr,Bauer:2001yt,Bauer:2002nz} has been developed and widely used until now. 

One another important ingredient for understanding lower energy physics is to properly separate soft and collinear physics. 
In SCET, although we distinguish collinear and soft parts at the operator level, even in actual calculation we need to subtract the overlapped contribution by hand for the complete factorization. For this, there has been introduced 'the zero-bin subtraction'~\cite{Manohar:2006nz}, where the overlapped soft contributions are subtracted in computing collinear parts.  
Usually, the collinear part has a larger renormalization scale or rapidity scale~\cite{Chiu:2011qc,Chiu:2012ir} than the soft part. So the subtraction of the soft contribution from the collinear part coincides with a typical matching procedure in the usual effective theory approach. 

Near threshold where collinear particles take most of the energy in collision, the remnant radiations are basically described as soft interactions. This soft part involves large logarithms at threshold that should be resummed to all orders in $\as$ for the precise estimation in the perturbative calculation. Hence it is important to properly obtain its renormalization behavior handling the threshold logarithms. However, the theoretical problem is that the threshold logarithms are also sensitive to collinear parts since the size of the logarithms is determined by the fact how much energy collinear particles take exactly.  

This problem can be resolved by introducing the collinear-soft (csoft) modes~\cite{Bauer:2011uc,Becher:2015hka,Chien:2015cka} additionally. As a result, in order to resum the threshold logarithms systematically, the original naive soft function needs to be refactorized into the global soft function and the csoft functions, where the global soft function can be obtained from the subtraction of the csoft contributions from the original soft function~\cite{Kidonakis:1998bk,Chay:2017bmy}. 

If we consider the dijet production near the threshold, the newly introduced csoft function can be combined with the collinear (jet) function, and this new combination forms the fragmentation function to a jet (FFJ)~~\cite{Kaufmann:2015hma,Kang:2016mcy,Dai:2016hzf,Dai:2017dpc}. Therefore, the invariant mass distribution for the dijet production is schematically given as  
\be 
\label{factsc}
\frac{d\sigma}{d\tau} \sim  \sigma_0 \cdot H(Q) \cdot S_{gs} \otimes D_{J_1/q} \otimes D_{J_2/\bar{q}} (\tau). 
\ee
Here $Q$ is the center of mass energy for $e^+e^-$-annihilation, and $\tau = M_{J_1J_2}^2/Q^2$ is close to 1 near threshold. $H$ and $S_{gs}$ are the hard and the global soft functions respectively, and $D_{J/q}$ is the FFJ initiated by quark. 
`$\otimes$' in Eq.~\eqref{factsc} denotes the convolution of $\tau$. If the production of dihadron is considered, one can immediately replace the FFJs with the (standard) fragmentation functions (FFs) for the hadrons such as $D_{h/q}$, while $H$ and $S_{gs}$ are universally given~\cite{Chay:2017bmy}.  

In this paper, using SCET we study the invariant mass distribution for the dijet production near threshold. 
In section~\ref{sectfact}, we derive the factorization theorem for the invariant mass distribution refactorizing the naive soft function into the global soft function and two csoft functions, and show the details of the factorization shown in Eq.~\eqref{factsc}. 
In section~\ref{resummation}, using the factorized result we resum the large threshold logarithms of $1-\tau$ to the accuracy of next-to-leading logarithms (NLL). We also compute the case of the heavy quark jet and compare its resummed results with the case of the light quark jet. Finally, in section~\ref{summary} we summarize. 

\section{Factorization theorem for the dijet invariant mass near threshold} 
\label{sectfact}

In the dijet limit, i.e., near threshold region of dijet, the produced jets from $e^+e^-$-annihilation move in opposite directions. In this case the scattering cross section with a total energy $Q$ can be factorized into hard, collinear, and soft parts, and it can be written as  
\begin{align}
\label{factHCS}
\sigma =& \sigma_0 H(Q,\mu) \mc{J}_{n}^q (E_{J_1}R,\mu) \mc{J}_{\n}^{\bar{q}} (E_{J_2}R,\mu) \nnb \\
&~~~\times \int dM^2 \frac{1}{N_c} \sum_{X_S} {\rm Tr} \langle 0 | Y_n^{\dagger} Y_{\n} |X_S \rangle \langle X_S |\delta \bigl((p_1+p_2)^2 - M^2\bigr) Y_{\n}^{\dg} Y_n | 0 \rangle,  
\end{align}
where $\sigma_0$ is the cross section at Born level, $H$ is the hard function, and $\mc{J}_{n}^q$ and $\mc{J}_{\n}^{\bar{q}}$ are the light quark jet functions in $n$- and $\n$-directions respectively. Here the lightcone vectors are given by $n^{\mu} = (1,\hat{\bl{n}})$ and $\n^{\mu}=(1,-\hat{\bl{n}})$, where $\hat{\bl{n}}$ is an unit vector. Also $p_1~(p_2)$ is the momentum of the jet in $n~(\n)$-direction. For clustering the jets, we employ inclusive $\mr{k_T}$-type algorithm~\cite{Catani:1993hr,Ellis:1993tq,Dokshitzer:1997in,Cacciari:2008gp}, where two particles are captured as a jet if they satisfy the following criterion, 
\be 
\theta < R, 
\ee 
where $\theta$ is the angle between two particles and $R$ is the jet radius. Throughout this paper, we consider a jet with small radius ($R\ll 1$). For simplicity, we set both the jets to have the same radius $R$. The energy of each jet can be considered to approximately have $E_J \approx Q/2$ near threshold. 

In Eq.~\eqref{factHCS}, $Y_n$ and $Y_{\n}$ are the soft Wilson lines decoupled from $n$- and $\n$-collinear field respectively. Since we consider the threshold region for the dijet production in $e^+e^-$-annihilation, the dijet momentum $p_1+p_2$ is very close to total momentum $q$ ($q^{\mu} = (Q,\bl{0}$)), and the remnant is given by the momentum of soft radiations from $Y_n$ and $Y_{\n}$. Therefore, the argument of the delta function in Eq.~\eqref{factHCS} becomes 
\begin{align} 
(p_1+p_2)^2 -M^2 &= (q-p_{X_S}^{\notin J})^2 - M^2 \nnb \\ 
\label{arg1}
&\approx Q^2 - M^2 - 2q\cdot p_{X_S}^{\notin J}=Q^2\bigl(1- \tau - \frac{2 p_{X_S}^{\notin J,0}}{Q}\bigr), 
\end{align}
where $\tau = M^2/Q^2$. $p_{X_S}^{\notin J}$ represents the soft momentum not to be included in the dijet, and reveals  nonlocality between the incoming and outgoing soft Wilson lines in Eq.~\eqref{factHCS}. Hence, when the final expression in Eq.~\eqref{arg1} is inserted in Eq.~\eqref{factHCS}, the zeroth component of the soft momentum, $p_{X_S}^{\notin J,0}$, can be expressed as a derivative operator taking the soft momentum from the soft Wilson lines. So the delta function in Eq.~\eqref{factHCS} should actually read 
\be
\delta \bigl((p_1+ p_2)^2 -M^2\bigr) \to \frac{1}{Q^2} \delta \bigl(1-\tau + \Theta_{\notin J}\frac{2i\partial^0}{Q}\bigr). \nnb 
\ee
Here `$\Theta_{\notin J}$' denotes the schematic expression that the derivative operator only acts on the soft states outside the dijet.  

Therefore, the cross section for $\tau$ can be written as 
\be 
\label{naivefact}
\frac{1}{\sigma_0}\frac{d\sigma}{d\tau} = H(Q,\mu) \mc{J}_{n}^q (E_JR,\mu) \mc{J}_{\n}^{\bar{q}} (E_JR,\mu) S (1-\tau,Q,R,\mu),
\ee
where the soft function $S$ is defined as 
\be
\label{naivesoft}
S (1-\tau,Q,R,\mu) = \frac{1}{N_c} \sum_{X_S} {\rm Tr} \langle 0 | Y_n^{\dagger} Y_{\n} |X_S \rangle \langle X_S |\delta \bigl(1-\tau + \Theta_{\notin J} \frac{2i\partial^0}{Q}\bigr) Y_{\n}^{\dg} Y_n | 0 \rangle.
\ee
At next-to-leading (NLO) in $\as$, the hard and the jet functions~\cite{Cheung:2009sg,Ellis:2010rwa,Chay:2015ila} are given by 
\begin{align}
H(Q,\mu) =& 1+ \frac{\as C_F}{2\pi}\Bigl(-3\ln\frac{\mu^2}{Q^2}-\ln^2\frac{\mu^2}{Q^2} - 8 + \frac{7\pi^2}{6} \Bigr)\ , \\
\label{lqjfNLO}
\mc{J}_n^q (E_JR,\mu)=&\mc{J}_{\n}^{\bar{q}} (E_JR,\mu) = 1+\frac{\as C_F}{2\pi} \Bigl(\frac{3}{2}\ln\frac{\mu^2}{E_J^2R^2}+\frac{1}{2}\ln^2\frac{\mu^2}{E_J^2R^2}+\frac{13}{2}-\frac{3\pi^2}{4} \Bigr)\ . 
\end{align}

If we compute radiative correction to the soft function $S(1-\tau,Q,R)$, we face the large logarithms not only with $1-\tau$ but also with small $R$. The presence of $\ln R$ can be inferred from the argument of the delta function in Eq.~\eqref{naivesoft}, where the nonzero value of $1-\tau$ can be only taken from the soft states out of the dijet, hence the radiative correction to the soft function $S(1-\tau,Q,R)$ can be sensitive to the jet boundary characterized by the jet radius $R$. 

In order to properly resum both the large logarithms with $1-\tau$ and $R$, we need to refactorize the soft function into the `global' soft function and the two `collinear-soft (csoft)' functions separating the full soft degrees of freedom into the global soft mode and the csoft modes~\cite{Bauer:2011uc,Becher:2015hka,Chien:2015cka,Chay:2017bmy}. Here the momenta of the refined modes scale as 
\begin{align}
\label{gssca}
p_{gs}^{\mu} &= (\n\cdot p_{gs},p_{gs}^{\perp}, n\cdot p_{gs}) \sim ~ Q(1-\tau, 1-\tau,1-\tau), \\  
\label{cssca}
p_{ncs}^{\mu} &\sim  E_J (1-\tau)(1, R,R^2),~~~p_{\n cs}^{\mu} \sim  E_J (1-\tau)(R^2, R,1),
\end{align}
where $E_J \sim Q/2$, and the subscript `{\it gs}' denotes the globl soft mode and `$ncs(\n cs)$' denotes $n(\n)$-csoft mode. 

Correspondently, the soft function can be refactorized as 
\begin{align}
\label{softrefac}
   S (1-\tau,Q,R,\mu) =& \int^1_\tau \frac{dz}{z} S_{gs} (1-z,Q,\mu) \\
   &\times \int^1_{\tau/z} \frac{dx}{x} S_{ncs}^q (1-x,E_JR,\mu)S_{\n cs}^{\bar{q}} (1-\frac{\tau}{xz},E_JR,\mu). 
   \nnb   
\end{align}
Here the refactorized soft functions are given as 
\begin{align}
\label{gsoft}
S_{gs} (1-z,Q,\mu) &= \frac{1}{N_c} \sum_{X_{gs}} {\rm Tr} \langle 0 | Y_n^{gs\dagger} Y_{\n}^{gs} |X_{gs} \rangle \langle X_{gs} |\delta \bigl(1-z +  \frac{2i\partial^0}{Q}\bigr) Y_{\n}^{gs\dg} Y_n^{gs} | 0 \rangle,  \\
\label{defSncs}
S_{ncs}^q (1-z,E_JR,\mu) &= \frac{1}{N_c} \sum_{X_{ncs}} {\rm Tr} \langle 0 | Y_n^{ncs\dagger} Y_{\n}^{ncs} |X_{ncs} \rangle \langle X_{ncs} |\delta \bigl(1-z + \Theta_{\notin J} \frac{i\n\cdot\partial}{Q}\bigr) Y_{\n}^{ncs\dg} Y_n^{ncs} | 0 \rangle. 
\end{align}
Also $S_{\n cs}^{\bar{q}} (1-z,E_JR)$ can be defined similarly with $S_{ncs}^q$ but employing $\n$-csoft mode. From the csoft momentum scaling in Eq.~\eqref{cssca}, the derivative operator for $S_{n cs}^q$ in Eq.~\eqref{defSncs} can be approximated as $2\partial^0 \approx \n\cdot \partial$. Similarly the derivative for $S_{\n cs}^{\bar{q}}$ becomes $2\partial^0 \approx n\cdot \partial$.

As seen in Eq.~\eqref{gssca}, the momentum scaling of the global soft mode is isotropic, so the mode cannot recognize the boundary of narrow jets characterized by small $R$ and the dependence on $R$ can be suppressed in $S_{gs}$. In computing radiative corrections to $S_{gs}$, we need to subtract the contributions of the $n$- and $\n$-csoft modes to avoid double counting~\cite{Chay:2017bmy,Kidonakis:1998bk,Hinderer:2014qta}. 
This subtraction can be also viewed as a matching procedure between the full soft function and the two csoft functions with the matching coefficient being given as $S_{gs}$. 
The NLO result of the global soft function is given by~\cite{Chay:2017bmy} 
\begin{align}
\label{SgsNLO}
    S_{gs}(Q(1-z))=\delta(1-z) + \frac{\as C_F}{\pi} \Biggl[\delta(1-z) \Bigl(\frac{1}{2} \ln\frac{\mu^2}{Q^2} -\frac{\pi^2}{4} \Bigr) - \Bigl(\frac{2}{1-z}\ln\frac{\mu^2}{Q^2(1-z)^2}\Bigr)_+\Biggr]\ , 
\end{align}
where the subscript `$+$' in the last term represents the plus function.  

The NLO result of the csoft function has been calculated in Ref.~\cite{Dai:2017dpc}, and it is given as 
\begin{align}
\label{ScsNLO}
   & S_{ncs}^q (1-z,E_JR) =S_{\n cs}^{\bar{q}}(1-z,E_JR) \\
    &~~~
    = \delta(1-z) 
    + \frac{\as C_F}{2\pi} \Biggl[- \delta(1-z) \Bigl(\frac{1}{2} \ln\frac{\mu^2}{E_J^2R^2} -\frac{\pi^2}{12} \Bigr) + \Bigl(\frac{2}{1-z}\ln\frac{\mu^2}{E_J^2R^2(1-z)^2}\Bigr)_+\Biggr]\ . \nnb 
\end{align}
Interestingly, when the csoft function $S_{ncs}^q$ is combined with the jet function $\mc{J}_{n}^q$, this combination become the fragmentation function to a jet (FFJ) in large $z$ region~\cite{Dai:2017dpc}, 
\be 
\label{refacFFJ}
 D_{J/q} (z\to 1,\mu;E_JR) = \mc{J}_{n}^q (E_JR,\mu) S_{ncs}^q(1-z,E_JR,\mu). 
\ee
Accordingly, the factorization theorem for Eq.~\eqref{naivefact} leads to 
\begin{align}
\label{factdijet}
    \frac{1}{\sigma_0}\frac{d\sigma}{d\tau} =& H(Q,\mu) \int^1_\tau \frac{dz}{z} S_{gs} (Q(1-z),\mu) \\
    &\times \int^1_{\tau/z} \frac{dx}{x} D_{J_1/q} (x,\mu;E_JR) D_{J_2/\bar{q}} (\frac{\tau}{xz},\mu;E_JR). \nnb
\end{align}
This result can be also applicable to the dihadron production near threshold. In this case each FFJ is replaced with the fragmentation functions to a hadron such as $D_{J_1/q}\to D_{h_1/q}$, while the hard function $H$ and the global soft function $S_{gs}$ remain unchanged~\cite{Chay:2017bmy}.   

Also the factorization theorem, Eq.~\eqref{factdijet}, can be applied to the heavy quark (HQ) dijet production based on the process $e^+e^-\to \mQ\bar{\mQ}$ near threshold as far as the heavy quark mass $m$ is given to be much smaller than $E_J~(\sim Q/2)$. The HQ FFJ to be employed in this case has been studied 
in Refs.~\cite{Dai:2018ywt,Dai:2021mxb}. Near threshold, similarly with Eq.~\eqref{refacFFJ}, the HQ FFJ can be refactorized as 
\be 
\label{refacHQFFJ}
 D_{J/Q} (z\to 1,\mu;E_JR,m) = \mc{J}_{n}^{\mQ} (E_JR,m,\mu) S_{ncs}^{\mQ}( 1-z,E_JR,m,\mu).   
\ee
Here $S_{ncs}^{\mQ}$ is introduced through the boosted heavy quark effective theory (bHQET) after integrating out degrees of freedom with offshellness $p^2 \sim m^2$~\cite{Dai:2021mxb,Kim:2020dgu}. 

The NLO results of the factorized functions are given by 
\begin{align}
\label{HQjfNLO}
    \mc{J}_{n}^Q (E_JR,m) &= 1+  \frac{\as C_F}{2\pi} \Biggl[\frac{3+b}{2(1+b)} \ln\frac{\mu^2}{E_J^2 R^2+m^2}+\frac{1}{2} \ln^2 \frac{\mu^2}{E_J^2 R^2+m^2}  \\
&~~+ \frac{2+\ln(1+b)}{1+b}  -\frac{1}{2} \ln^2 (1+b) + f(b)+g(b)-\mr{Li}_2 (-b) + 2 -\frac{\pi^2}{12} \Biggr]\ .\nnb \\
S_{ncs}^{\mQ}( 1-z,E_JR,m) &= \delta(1-z) 
+\frac{\as C_F}{2\pi} \Biggl\{\delta(1-z) \Bigl[
\frac{b}{1+b}\ln \frac{\mu^2}{E_J^2 R^2+m^2}-\frac{1}{2} \ln^2 \frac{\mu^2}{E_J^2 R^2+m^2}
\nnb \\
&-\frac{1}{1+b} \ln (1+b) +\frac{1}{2}\ln^2(1+b)+\mr{Li}_2 (-b)+\frac{\pi^2}{12}\Bigr] \nnb \\
\label{SJhNLO}
&+\Bigl[\frac{2}{1-z}\bigl(\ln\frac{\mu^2}{(E_J^2 R^2+m^2)(1-z)^2}-\frac{b}{1+b}\bigr)\Bigr]_+ \Biggr\}\ , 
\end{align}
where $b\equiv m^2/E_J^2R^2$, and the functions $f(b)$ and $g(b)$ can be given as the integral forms, 
\bea
\label{fb}
f(b) &=& \int^1_0 dz \frac{1+z^2}{1-z} \ln\frac{z^2+b}{1+b}, \\
\label{gb}
g(b) &=& \int^1_0 dz \frac{2z}{1-z}\Bigl(\frac{1}{1+b}-\frac{z^2}{z^2+b}\Bigr) 
\eea
Note that, if we take the limit $m\to 0$, the NLO results for $\mc{J}_{n}^Q$ and $S_{ncs}^{\mQ}$ in Eqs.~\eqref{HQjfNLO} and \eqref{SJhNLO} recover the results for a light quark, Eqs.~\eqref{lqjfNLO} and \eqref{ScsNLO} respectively. 

\section{Resummation of large logarithms}
\label{resummation}

In order to resum the large logarithms, we consider the dijet invariant mass distribution in the moments space, where the factorization is expressed as a multiplication of factorized functions rather than convolution. The $N$th-moments are given as 
\begin{align}
\label{momdijet}
   \tilde{\sigma}_N\equiv\frac{1}{\sigma_0} \int^1_0 d\tau \tau^{-1+N} \Bigl(\frac{d\sigma}{d\tau}\Bigr) &= H(Q,\mu) \mc{J}_{n}^f (E_JR,m_f,\mu) \mc{J}_{\n}^{\bar{f}} (E_JR,m_f,\mu) \\
   &\times \tilde{S}_{gs} (Q,\bar{N},\mu) \tilde{S}_{ncs}^f (E_JR,\bar{N},m_f,\mu)  \tilde{S}_{\n cs}^{\bar{f}} (E_JR,\bar{N},m_f,\mu), \nnb 
\end{align}
where $f=q,\mQ$, and we set $m_q=0$ and $m_{\mQ} = m$. Here, at NLO in $\as$, the moments of the soft and the csoft functions in large $N$ limit are given as 
\begin{align}
\label{tSgs}
    \tilde{S}_{gs} (Q/\bar{N}) &= 1 + \frac{\as C_F}{2\pi} \Bigl(\ln^2 \frac{\mu^2\bar{N}^2}{Q^2} + \frac{\pi^2}{6} \Bigr), \\
\label{tSncsq}
    \tilde{S}_{ncs}^q (E_JR,\bar{N}) &=\tilde{S}_{\n cs}^{\bar{q}} (E_JR,\bar{N}) = 1 + \frac{\as C_F}{2\pi} \Bigl(-\frac{1}{2} \ln^2 \frac{\mu^2\bar{N}^2}{E_J^2R^2} - \frac{\pi^2}{4} \Bigr), \\
    \tilde{S}_{ncs}^{\mQ} (E_JR,\bar{N},m) &=\tilde{S}_{\n cs}^{\bar{\mQ}} (E_JR,\bar{N},m) = 
    1+ \frac{\as C_F}{2\pi}\Bigl[\frac{b}{1+b} \ln \frac{\mu^2\bar{N}^2}{B^2}-\frac{1}{2}\ln^2\frac{\mu^2\bar{N}^2}{B^2}-\frac{\pi^2}{4} \nnb \\
\label{tSncsQ}
&\hspace{3cm}-\frac{1}{1+b} \ln (1+b) +\frac{1}{2}\ln^2(1+b)+\mr{Li}_2 (-b)\Bigr],
\end{align}
where $\bar{N} \equiv N e^{\gamma_E}$, $b \equiv m^2/(E_JR)^2$, and $B^2 \equiv E_J^2R^2+m^2$. 

At the accuracy of next-to-leading logarithms (NLL), the anomalous dimensions satisfying the RG equations, $df/d\ln\mu = \gamma_f \cdot f$, are given by 
\begin{align}
    \gamma_H =& -2\Gamma_C \ln \frac{\mu^2}{Q^2}+\frac{\as C_F}{2\pi} (-6), \\
    \gamma_{\mc{J}}^q =& \Gamma_C \ln \frac{\mu^2}{E_J^2R^2}+\frac{\as C_F}{2\pi}\cdot 3, \\
    \gamma_{\mc{J}}^{\mQ} =& \Gamma_C \ln \frac{\mu^2}{B^2}+\frac{\as C_F}{2\pi}\frac{3+b}{1+b}\ ,  
\end{align}
where $\gamma_{\mc{J}}^{q}~(\gamma_{\mc{J}}^{\mQ})$ is the anomalous dimension for $\mc{J}^{q}_{n,\n}~(\mc{J}^{\mQ}_{n,\n})$. And $\Gamma_C$ is the cusp anomalous dimension~\cite{Korchemsky:1987wg,Korchemskaya:1992je}. When it is expanded as $\Gamma_{C} = \sum_{k=0} \Gamma_{k}\cdot (\as/4\pi)^{k+1}$, we need first two coefficients,
\be
\Gamma_{0} = 4C_F,~~~\Gamma_{1} = 4C_F \Bigl[\bigl(\frac{67}{9}-\frac{\pi^2}{3}\bigr) C_A - \frac{10}{9} n_f\Bigr].
\ee
Also the anomalous dimensions for the global and csoft functions in the moments space are given as 
\begin{align}
    \gamma_{gs} &=2\Gamma_C \ln \frac{\mu^2 \bar{N}^2}{Q^2}, \\
    \gamma_{cs}^q &= -\Gamma_C \ln \frac{\mu^2 \bar{N}^2}{E_J^2R^2},~~~\gamma_{cs}^{\mQ} = -\Gamma_C \ln \frac{\mu^2 \bar{N}^2}{B^2}+\frac{\as C_F}{2\pi}\frac{2b}{1+b}\ , 
\end{align}
where $\gamma_{cs}^{q}~(\gamma_{cs}^{\mQ})$ is the anomalous dimension for $S^{q}_{n,\n cs}~(S^{\mQ}_{n,\n cs})$.

Through RG evolution using the anomalous dimensions, we resum large logarithms and exponentiate them to NLL accuracy. The result for the distribution of dijet with light quarks is given as  
\begin{align}
\label{resumL}
    \frac{1}{\sigma_0} \frac{d\sigma}{d\tau} &= H (Q,\mu_h) \bigl[\mc{J}_n^q (E_JR,\mu_j)\bigr]^2 \exp[\mc{M}_L (\mu_h,\mu_j,\mu_{gs},\mu_{cs})] \\ 
    &\times (1-\tau)^{-1+\eta}~\tilde{S}_{gs} \Bigl[\ln\frac{\mu_{gs}^2}{Q^2(1-\tau)^2} - 2\partial_{\eta}\Bigr]
    \tilde{S}_{cs}^2 \Bigl[\ln\frac{\mu_{cs}^2}{E_J^2R^2(1-\tau)^2} - 2\partial_{\eta}\Bigr]\frac{e^{-\gamma_E \eta}}{\Gamma(\eta)}. \nnb
\end{align}
Here, for simplicity, we have identified both the functions in $n$- and $\n$-directions such as  $\mc{J}_n^q =\mc{J}_{\n}^{\bar{q}}$ and $\tilde{S}_{cs} \equiv \tilde{S}_{ncs}^q = \tilde{S}_{\n cs}^{\bar{q}}$.
And $\mu_i~(i=h,j,gs,cs)$ are the characteristic scales to minimize the large logarithms in the factorized functions. They are roughly given by  
\be
\label{char1}
\mu_h \sim Q,~~\mu_j \sim E_JR,~~\mu_{gs} \sim Q(1-\tau),~~\mu_{cs} \sim E_JR(1-\tau).
\ee 
In Eq.~\eqref{resumL}, the global-soft and the csoft functions are basically the same as Eqs.~\eqref{tSgs} and \eqref{tSncsq}, but their logarithmic terms have been replaced with ones to include $-2\partial_{\eta}$. 

After RG evolution of each factorized function from a factorization scale to the characteristic scale, we obtain the exponentiation factor $\mc{M}_L$ in Eq.~\eqref{resumL}. To NLL accuracy, it is given as  
\begin{align}
\label{expML}
    \mc{M}_L (\mu_h,\mu_j,\mu_{gs},\mu_{cs}) &= 4 S_{\Gamma} (\mu_h,\mu_{gs}) - 4 S_{\Gamma} (\mu_j,\mu_{cs})
    + 2 \ln\frac{\mu_h^2}{Q^2} a_{\Gamma} (\mu_h,\mu_{gs}) \\
    &-2\ln\frac{\mu_j^2}{E_J^2R^2} a_{\Gamma} (\mu_j,\mu_{cs}) 
    -\frac{6 C_F}{\beta_0}\ln\frac{\as(\mu_h)}{\as(\mu_j)}\ , \nnb
\end{align}
where $S_{\Gamma}$ and $a_{\Gamma}$ are the evolution factors with $\Gamma_C$, and they are defined as 
\be
S_{\Gamma} (\mu_1,\mu_2) = \int^{\alpha_1}_{\alpha_2} \frac{d\as}{b(\as)} \Gamma_{C}(\as) \int^{\as}_{\alpha_1}
\frac{d\as'}{b(\as')},~~~a_{\Gamma}(\mu_1,\mu_2) = \int^{\alpha_1}_{\alpha_2} \frac{d\as}{b(\as)} \Gamma_C(\as).
\ee
Here $\alpha_{1,2} \equiv \as (\mu_{1,2})$, and $b(\as)=d\as/d\ln\mu$ is the QCD beta function, which is expanded as $b(\as)=-2\as\sum_{k=0}\beta_k (\as/4\pi)^{k+1}$. The evolution parameter $\eta$ in Eq.~\eqref{resumL} is given by $\eta = 4a_{\Gamma}(\mu_{gs},\mu_{cs})$ and it is positive since $\mu_{gs} > \mu_{cs}$. 

When a heavy quark $(\mQ)$ initiates the jet, the exponentiation factor in eq.~\eqref{resumL} depends on the heavy quark mass $m$, and it is given as  
\begin{align}
\label{expMH}
    \mc{M}_H (\mu_h,\mu_j,\mu_{gs},\mu_{cs}) &= 4 S_{\Gamma} (\mu_h,\mu_{gs}) - 4 S_{\Gamma} (\mu_j,\mu_{cs})
    + 2 \ln\frac{\mu_h^2}{Q^2} a_{\Gamma} (\mu_h,\mu_{gs}) \\
    &-2\ln\frac{\mu_j^2}{B^2} a_{\Gamma} (\mu_j,\mu_{cs}) 
    -\frac{2C_F}{\beta_0}\Bigl(\frac{3+b}{1+b} \ln\frac{\as(\mu_h)}{\as(\mu_j)}+\frac{2b}{1+b} \ln\frac{\as(\mu_h)}{\as(\mu_{cs})}\Bigr)\ , \nnb
\end{align}
where the characteristic scales for $\mc{J}_{n,\bar{n}}^{\mQ}$ and $\tilde{S}_{ncs,\bar{n} cs}^{\mQ}$ are respectively given by 
\be
\label{char2}
\mu_j \sim B,~~\mu_{cs} \sim B(1-\tau), 
\ee
where $B = \sqrt{E_J^2R^2+m^2}$. Correspondently, the logarithm for $\tilde{S}_{cs}$
in Eq.~~\eqref{resumL} should be replaced with $\ln (\mu_{cs}^2/B^2(1-\tau)^2)$ in the case of the heavy quark.  

At the higher order in $\as$, i.e., starting from two loop order, large nonglobal logarithms (NGLs)~\cite{Dasgupta:2001sh,Banfi:2002hw} arise in the factorization between the collinear modes  ($\mc{J}_{n(\n)}^f$) and the csoft modes ($\tilde{S}_{n(\n) cs}$), where both the modes can recognize the jet boundary. These NGLs can contribute at the accuracy of NLL. 
In case of the jet production with a light quark initiation, we use the resummed result of leading NGLs in the large $N_c$ limit obtained in Ref~~\cite{Dasgupta:2001sh}, and the contribution to the dijet production is expressed as  
\be\label{RNGL}
\Delta^q_{\mr{NG}} (\mu_{c},\mu_{cs}) = \exp\Biggl(-2 C_A C_F \frac{\pi^2}{3} \Bigl(\frac{1+(at)^2}{1+(bt)^c}\Bigr) t^2 \Biggr)\ ,
\ee
where 
\be
t=\frac{1}{\beta_0} \ln{\frac{\as(\mu_{cs})}{\as(\mu_{c})}} \sim -\frac{1}{\beta_0}\ln \Bigl(1-\frac{\beta_0}{4\pi} \as(\mu_c) \ln \frac{\mu_c^2}{\mu_{cs}^2}\Bigr)\ .
\ee 
Here the fit parameters given as $a=0.85 C_A,~b=0.86C_A$, and $c=1.33$~\cite{Dasgupta:2001sh}.
Resummation of NGLs for the heavy quark jet production is beyond the scope of this paper. For a recent study of this case we refer to Ref.~\cite{Balsiger:2020ogy}. 

\begin{figure}[h]
     \centering
     \subfloat[][Light quark dijet production]{\includegraphics[width=0.5\textwidth]{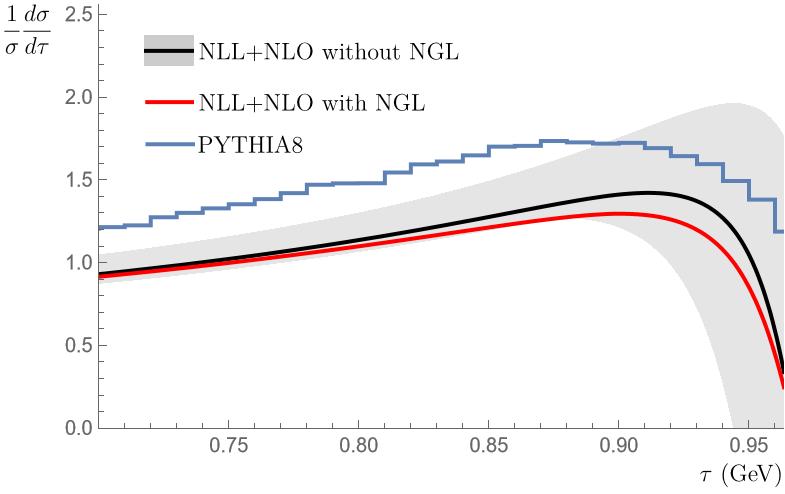}\label{lmzfig}} 
     \subfloat[][Heavy quark dijet production]{\includegraphics[width=0.5\textwidth]{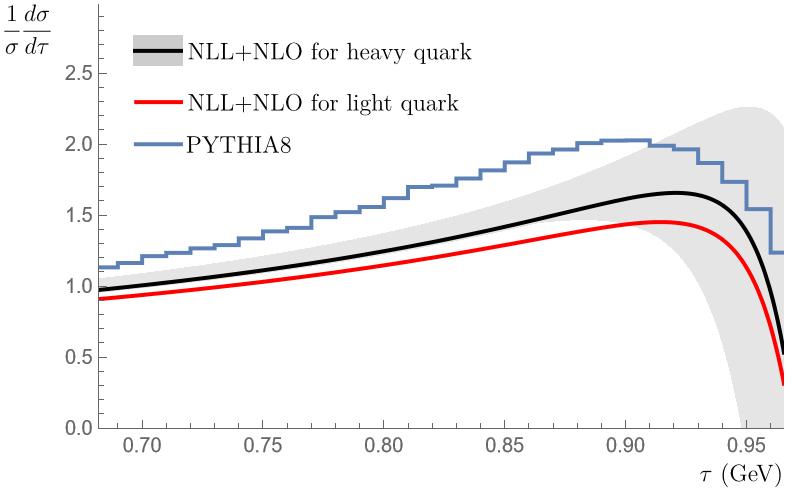}\label{lhmzfig}}
     \caption{Dijet invariant mass distributions at $Q = m_Z$ in $e^+e^-$-annihilation.}
     \label{mz}
\end{figure}

\begin{figure}[h]
     \centering
     \subfloat[][Light quark dijet production]{\includegraphics[width=0.5\textwidth]{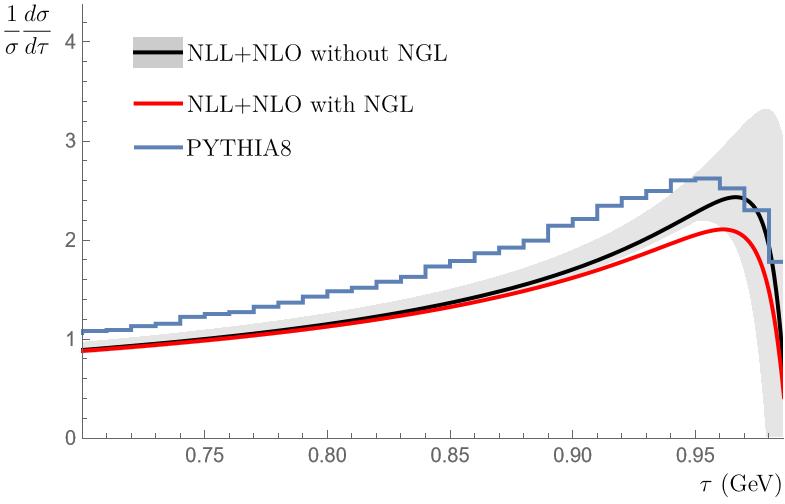}\label{l240fig}}
     \subfloat[][Heavy quark jets production]{\includegraphics[width=0.5\textwidth]{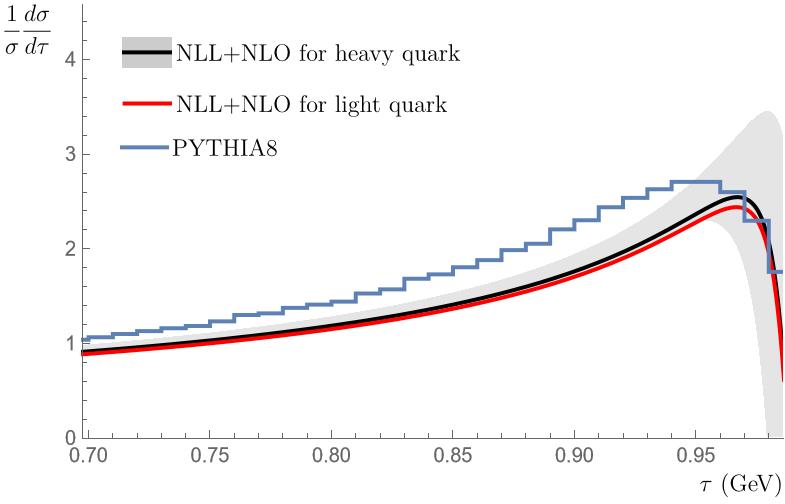}\label{l240fig}}
     \caption{Dijet invariant mass distributions at $Q = 240~\rm GeV$ in $e^+e^-$-annihilation.}
     \label{240}
\end{figure}

We numerically estimate the differential cross section for $\tau$ shown in Eqs.~\eqref{resumL} at the accuracy of NLL+NLO, where we have included the NLO corrections to each factorized function at the fixed order in $\as$. 
We also compare these theoretical results with dijet events simulated through PYTHIA~8.3~\cite{Bierlich:2022pfr}. We set the jet radius as $R=0.3$. In using PYTHIA~8.3, in order to focus the events in the dijet limit, we employ the jet energy veto $E_{veto} \sim E_J R$, while this choice of the veto does not have effects on our theoretical calculation near threshold. 
In FIG.~\ref{mz} we consider the dijet invariant mass distribution at $Z$-pole ($m_Z = 91.2~\rm GeV)$ and we do at CM energy $Q=240~\rm GeV$ in FIG.~\ref{240}. In both FIGs, the left columns~(a) describe the light quark dijet production, and the right columns the $b$-quark dijet production. 

In each figure, the black solid line represents our default calculation (without the NGL effects) using the characteristic scales shown in Eqs.~\eqref{char1} and \eqref{char2}. And the gray band is the error estimation varying characteristic scales $\mu_i$ form $\mu_i/2$ to $2\mu_i$. 
As $Q$ becomes larger, the shape of the distributions becomes narrower, and the peak position becomes close to $\tau = 1$. 
As seen in FIG.~\ref{mz}-(b), the $b$-dijet production gives a sizable enhancement when compared with the case of the light quark, which results from the fact the heavy quark mass plays a role in widening the jet size. 
In the left sides of FIGs.~\ref{mz} and \ref{240}, the red lines denote the distributions including the NGL effects based on the estimation, Eq.~\eqref{RNGL}. 
When compared with our default calculations, the NGL effects give some slight suppression around the peak position.


\section{Summary} 
\label{summary}

In this paper, to resum the large logarithms near threshold in dijet production, we have derived the factorization theorem, where the original soft function that is responsible for the threshold logarithms can be refactorized into the global soft function and the two csoft functions as shown in Eq.~\eqref{softrefac}. 
The refactorized csoft function can be combined with the collinear jet function to form the FFJ as shown in Eqs.~\eqref{refacFFJ}. 

The factorization theorem derived here can be easily applicable to other processes near threshold. When we consider the heavy quark dijet production, we can employ the HQ FFJ as seen in Eq.~\eqref{refacHQFFJ}. If we consider hadrons in the final states, we can use the hadron's FFs instead of the FFJs.\footnote{\baselineskip 3.0ex
Interestingly, like the FFJ, the FF to a heavy hadron can be additionally factorized as the collinear~($\mu_c \sim m$) and the csoft~($\mu_{cs} \sim m(1-z)$) functions~\cite{Neubert:2007je,Fickinger:2016rfd}
} 
Also the global soft function can be commonly presented in other processes such as dihadron production or Drell-Yan production.\footnote{\baselineskip 3.0ex
If we consider the dijet production at the LHC, the relevant global soft function can be obtained by subtracting the csoft contributions to the parton distribution functions as well as to the FFJs in the final states~\cite{Kidonakis:1998bk}.}

Finally, in FIGs.~\ref{mz} and \ref{240}, we have shown the resummed results for the dijet invariant mass to the NLL accuracy, and compared both the cases of the light quark jets and the heavy quark jets. More precise resummed results and their comparison with experimental data will be a cornerstone for understanding perturbative and nonperturbative aspects in QCD and discovering new physics in future colliders.

\acknowledgments

This study was supported by the Research Program funded by Seoul National University of Science and Technology.


\bibliographystyle{apsrev4-2}
\bibliography{fullrefs}


\end{document}